\documentclass[11pt]{article}

\usepackage{amssymb}
\usepackage{amsmath}
\usepackage{amsfonts}

\newcommand{\R}{\mathbb{R}}

\newcommand{\be}{\begin{equation}}
\newcommand{\ee}{\end{equation}}
\newcommand{\bea}{\begin{eqnarray}}
\newcommand{\eea}{\end{eqnarray}}

\newcommand{\ed}{\end{document}}

\newcommand{\LL}{{\cal L}}
\newcommand{\bL}{{\bf L}}
\newcommand{\ro}{\partial}

\newcommand{\eq}[1]{Eq.~(\ref{#1})}
\newcommand{\eqs}[1]{Eqs.~(\ref{#1})}
\newcommand{\e}[1]{(\ref{#1})}
\begin{document}

\title{Intertwined  Hamiltonians in Two Dimensional Curved Spaces}
\author{Keivan Aghababaei Samani\thanks{E-mail address:
samani@cc.iut.ac.ir}~ and Mina Zarei \\ \\
{\it Department of Physics, Isfahan University of Technology (IUT),}\\
{\it Isfahan 84156, Iran}}
\date{ }
\maketitle
\begin{abstract}
The problem of intertwined Hamiltonians in two dimensional curved
spaces is investigated. Explicit results are obtained  for
Euclidean plane,Minkowski plane, Poincar{\' e} half plane
($AdS_2$), de Sitter Plane ($dS_2$), sphere, and torus. It is
shown that the intertwining operator is related to the Killing
vector fields and the isometry group of corresponding space. It
is shown that the intertwined potentials are closely connected to
the integral curves of the Killing vector fields. Two problems of
considered as applications of the formalism presented in the
paper. The first one is the problem of Hamiltonians with
equispaced energy levels and the second one is the problem of
Hamiltonians  whose spectrum are like the spectrum of a free
particle.
\end{abstract}
%\vspace{2mm}
%PACS numbers: 03.65.Bz\\
%\vspace{2mm}

%\baselineskip=24pt
\newpage

\section{Introduction}
The {\it Intertwining} relationship between differential
operators is widely studied in recent years~\cite{1, 2, 3}. This
is simply the following relationship between three differential
operators $H_1$, $H_2$ and $\LL$
    \be
    \LL H_1=H_2 \LL\;.
    \label{first}
    \ee
We call $H_1$ and $H_2$  the intertwined Hamiltonians and $\LL$
the intertwining operator. One also calls \eq{first} a {\it
Darboux Transformation}. In this case we say that $H_2$ is
Darboux transform of $H_1$ by $\LL$. Intertwined Hamiltonians are
essentially isospectral: if $\psi$ is an eigenfunction of $H_1$
with eigenvalue $E\ne 0$, then $\LL \psi$ is an eigenfunction of
$H_2$ with the same eigenvalue. When $H_1$ and $H_2$ are
Hermitian, \eq{first} immediately results in
    \be
    [H_1, \LL^\dagger \LL]=[H_2,\LL \LL^\dagger]=0\;.
    \label{symm}
    \ee
Therefore $\LL^\dagger \LL$ and $\LL \LL^\dagger$ are dynamical
symmetries of $H_1$ and $H_2$ respectively. It is worth noting
that the above mentioned properties are in general regardless of
dimension and form of Hamiltonians.

\eq{first} appears in various fields in Physics and Mathematics.
In supersymmetric Quantum Mechanics the intertwining relation is
used to find exact solutions of Schr\"odinger equation for shape
invariant potentials and to construct new exactly solvable
potentials from the known ones~\cite{rev}. It is also appears in
the algebraic relations of topological symmetries which are some
generalizations of supersymmetry~\cite{first,npb,stat}.

In AdS/CFT correspondence, intertwining relations are used to
realize the equivalence between the representations describing
the bulk fields and the boundary fields~\cite{dobrev}.

As \eq{first} preserves many spectral properties of the
Hamiltonian, it has been widely used in the investigation of
soliton equations~\cite{soliton} and in the study of bispectral
property~\cite{bispectral}.

In this paper we investigate the general solutions of \eq{first}
in a 2D curved space. Classical and Quantum dynamics on a 2D
curved space is widely studied~\cite{chaos, gutz}. In this paper
we study properties of intertwined Hamiltonians in such spaces.
The complete exact solutions are obtained for Euclidean plane,
Minkowski plane, Poincar\`e half plane, de Sitter plane, Sphere,
and Torus. It is shown that the solutions are related to Killing
vector fields and isometry group of corresponding space. Our
solutions are also closely related to the problem of
superintegrability in 2D spaces~\cite{1,2, winter, kalnins}.

Consider a two dimensional curved surface with coordinates $x^a$,
$a=1,2$ and metric $g_{ab}$. Suppose that $H_1$ and $H_2$ are two
Hamiltonians defined as below
    \bea
    &&H_1=-\nabla^2+V_1(x^1,x^2)\;, \label{h1}\\
    &&H_2=-\nabla^2+V_2(x^1,x^2)\;, \label{h2}
    \eea
where $\nabla^2$ is the Laplasian operator,
$\nabla^2=\frac{1}{{\sqrt g}}\ro_a\left({\sqrt
g}g^{ab}\ro_b\right)$, and $g=\det (g_{ab})$ and $g^{ab}$ is the
inverse of $g_{ab}$, i.e. $g^{ac}g_{cb}=\delta^a_b$. We call
$V_1$ and $V_2$ intertwined potentials if \eq{first} holds. In
this paper we will take $\LL$  to be a first order differential
operator,
    \be
    \LL=L_0+L^a\ro_a\equiv L_0+\bL \;,
    \label{L}
    \ee
where $L_0$ and  $L^a$, $a=1,2$,  are real valued functions of
$x^1$ and $x^2$. Our aim is to find $L_0$, $L^a$, $V_1$, and $V_2$
such that \eq{first} holds. Substituting  $\LL$, $H_1$ and $H_2$
from \eqs{L},  \e{h1}, and \e{h2} into \eq{first}, we arrive at
the following equations
    \bea
    &&L^c\ro_cg^{ab}-g^{bc}\ro_cL^a-g^{ac}\ro_cL^b=0\;,a,b=1,2\;, \label{e2}\\
%    &&PL^a=\nabla^2L^a+2g^{ab}\ro_bL_0-L^c\ro_c\left(\frac{1}{{\sqrt
%    g}}\ro_b\left({\sqrt g}g^{ab}\right)\right)\;,a=1,2 \label{e1}\\
    &&PL^a=2g^{ab}\ro_bL_0\;,a=1,2 \label{e1}\\
    &&PL_0=\nabla^2L_0+L^c\ro_cV_1\;,\label{e0}
    \eea
where $P=V_2-V_1$. \eqs{e2} -- \e{e0} are completely general, i.e.
they are correct in all dimensions. In this paper we will
investigate only two dimensional spaces. \eq{e2} is the well
known Killing equation. Its solutions are the Killing vector
fields of the corresponding space. The rest of the paper is
devoted to solve \eqs{e2} -- \e{e0} in some two dimensional
spaces, namely Euclidean plane, Minkowski plane, Poincar\`{e} half
plane, de Sitter plane, sphere, and torus. Finally we give two
examples as applications of the formalism presented in this
paper. The first one is the problem of Hamiltonians with
equispaced energy levels on two dimensional surfaces and the
second one is about the particles which move in a nontrivial
potential, but their spectrum are like free particle spectrum. We
call these particles {\it free like} particles.

\section{Euclidean Plane}
Euclidean plane the simplest case in which the metric is
$g_{ab}=diag(1,1)$. We use the notation $x^1=x$, $x^2=y$. Then
\eq{e2} takes the following form
    \bea
    &&\ro_x L^x=0\;,\label{u1}\\
    &&\ro_y L^y=0\;,\label{u2}\\
    &&\ro_x L^y+\ro_y L^x=0\;, \label{u3}
    \eea
The above equations has a general solution, ${\bf
L}=L^x\ro_x+L^y\ro_y$, of the form
    \be
    \bL = \alpha \bL_1+\beta \bL_2+\gamma \bL_3 \;, \label{l}
    \ee
where $\alpha$, $\beta$ and $\gamma$ are real constants and
    \bea
    &&\bL_1=\ro_x\;,\label{lu1}\\
    &&\bL_2=\ro_y\;, \label{lu2}\\
    &&\bL_3=y\ro_x-x\ro_y\;. \label{lu3}
    \eea
It is easily seen that $\bL_1$, $\bL_2$ and $\bL_3$ satisfy the
algebra of the isometry group of the Euclidean plane, namely
$E(2)$,
    \be
    [\bL_1,\bL_2]=0,\;\;[\bL_2,\bL_3]=\bL_1,\;\;[\bL_3,\bL_1]=\bL_2\;,
    \label{e2algebra}
    \ee
As the general solution for $\bL$ is a combination of $\bL_1$,
$\bL_2$ and $\bL_3$, we classify the solutions in three classes
as follows. It is worth noting that from a Physical (or
geometrical) point of view these classes are not independent. For
example class 1 and class 2 below  are equivalent. Therefore this
classification is from an algebraic point of view which means how
one can construct the most general intertwining operator.
\subsection{Class 1}
In this class we take $\bL=\bL_1 \equiv \ro_x$. taking  this solution for $\bL$, one
can easily see that \eq{e1} leads to
    \bea
    && P=2\ro_x L_0\;,\label{l0u1}\\
    &&\ro_y L_0=0\;, \label{l0u2}
    \eea
\eq{l0u2} means that $L_0$ is a function of $x$ only, therefore if we choose $L_0$
arbitrarily, we can find $P$ easily:
    \bea
    &&L_0=L_0(x)\;, \label{l0u}\\
    &&P=2 L'_0\label{pu}
    \eea
where  $L'_0(x)=dL_0/dx$ (Throughout this paper a `$\prime$' sign
means the derivative of a function with respect to it's argument).

Next we use  \eq{e0}  to find $V_1$. Inserting the above results in \eq{e0} we get
    \be
    2L_0 L'_0=L_0{''}+\ro_x V_1\;. \label{v1ue}
    \ee
This equation is easily integrated and one arrives at the following general solution
for $V_1$ and $V_2$.
    \bea
    &&V_1(x,y)=L_0^2(x)-L'_0(x)+f(y)\;,\label{v1u}\\
    &&V_2(x,y)=L_0^2(x)+L'_0(x)+f(y)\;,\label{v2u}
    \eea
where $f(y)$ is an arbitrary  function of $y$.
\subsection{Class 2}
In this class we take $\bL=\bL_2=\ro_y$. This class is similar to class 1. The
solutions can be obtained  easily from the class 1 solutions by interchanging the
role of $x$ and $y$
    \bea
    &&L_0=L_0(y)\;, \label{l0uc2}\\
    &&P=2 L'_0\label{puc2}\\
    &&V_1(x,y)=L_0^2(y)-L'_0(y)+f(x)\;,\label{v1uc2}\\
    &&V_2(x,y)=L_0^2(y)+L'_0(y)+f(x)\;.\label{v2uc2}
    \eea
\subsection{Class 3}
In this class we take $\bL=\bL_3 \equiv y\ro_x-x \ro_y$. With this choice for $\bL$,
\eqs{e1} and \e{e0} take the following form
    \bea
    &&Py=2\ro_x L_0\;, \label{eu1}\\
    &&-Px=2\ro_y L_0\label{eu2}\\
    &&P L_0=\ro_x^2 L_0 +\ro_y^2 L_0+y\ro_x V_1 -x\ro_y V_1\;,\label{eu3}
    \eea
These equations are easily solved with the following change of variables
    \be
    u=\frac{x}{y}\;,v=x^2+y^2\;. \label{changeu3}
    \ee
The  general solutions are
    \bea
    &&L_0=L_0(u)\;, \label{l0uc3}\\
    &&P=\frac{2}{y^2} L'_0\label{puc3}\\
    &&V_1(u,v)=\frac{1}{v}\left(L_0^2(u)-(1+u^2)L'_0(u)+f(v)\right)\;,\label{v1uc3}\\
    &&V_2(u,v)=\frac{1}{v}\left(L_0^2(u)+(1+u^2)L'_0(u)+f(v)\right)\;.\label{v2uc3}
    \eea
The meaning of the new variables $u$ and $v$ is worth noting. The family of curves
given by $v=const$ is the family of integral curves of the Killing vector field
$\bL=y\ro_x-x \ro_y$. Also the curves given by $u=const$ are normal to the curves
given by $v=const$.
\section{Minkowski plane}
For Minkowski plane  the metric is $g_{ab}=diag(-1,1)$. We use the
notation $x^1=t$, $x^2=x$. Then \eq{e2} takes the following form
    \bea
    &&\ro_t L^t=0\;,\label{m1}\\
    &&\ro_x L^x=0\;,\label{m2}\\
    &&\ro_t L^x-\ro_x L^t=0\;, \label{m3}
    \eea
The above equations has a general solution, ${\bf
L}=L^t\ro_t+L^x\ro_x$, of the form
    \be
    \bL = \alpha \bL_1+\beta \bL_2+\gamma \bL_3 \;, \label{lm}
    \ee
where $\alpha$, $\beta$ and $\gamma$ are real constants and
    \bea
    &&\bL_1=\ro_t\;,\label{lm1}\\
    &&\bL_2=\ro_x\;, \label{lm2}\\
    &&\bL_3=x\ro_t+t\ro_x\;. \label{lm3}
    \eea
It is easily seen that $\bL_1$, $\bL_2$ and $\bL_3$ satisfy the
algebra of the isometry group of the Minkowski  plane, Which is a
subgroup of Poincar{\' e} group
    \be
    [\bL_1,\bL_2]=0,\;\;[\bL_2,\bL_3]=\bL_1,\;\;[\bL_1,\bL_3]=\bL_2\;,
    \label{pm2algebra}
    \ee
As the general solution for $\bL$ is a combination of $\bL_1$,
$\bL_2$ and $\bL_3$, we classify the solutions in three classes
as follows
\subsection{Class 1}
In this class we take $\bL=\bL_1 \equiv \ro_t$. taking  this
solution for $\bL$, one can easily see that \eq{e1} leads to
    \bea
    && P=-2\ro_t L_0\;,\label{l0m1}\\
    &&\ro_x L_0=0\;, \label{l0m2}
    \eea
\eq{l0m2} means that $L_0$ is a function of $t$ only, therefore if
we choose $L_0$ arbitrarily, we can find $P$ easily:
    \bea
    &&L_0=L_0(t)\;, \label{l0m}\\
    &&P=-2 L'_0\label{pm}
    \eea
where  $L'_0(t)=dL_0/dt$.

Next we use  \eq{e0}  to find $V_1$. Inserting the above results
in \eq{e0} we get
    \be
    2L_0 L'_0=L_0{''}+\ro_t V_1\;. \label{v1me}
    \ee
This equation is easily integrated and one arrives at the
following general solution for $V_1$ and $V_2$.
    \bea
    &&V_1(t,x)=-L_0^2(t)+L'_0(t)+f(x)\;,\label{v1m}\\
    &&V_2(t,x)=-L_0^2(t)-L'_0(t)+f(x)\;,\label{v2m}
    \eea
where $f(x)$ is an arbitrary  function of $x$.
\subsection{Class 2}
In this class we take $\bL=\bL_2=\ro_x$. This class is very
similar to class 1. The solutions can be obtained  easily in the
same manner
    \bea
    &&L_0=L_0(x)\;, \label{l0mc2}\\
    &&P=2 L'_0\label{pmc2}\\
    &&V_1(t,x)=L_0^2(x)-L'_0(x)+f(t)\;,\label{v1mc2}\\
    &&V_2(t,x)=L_0^2(x)+L'_0(x)+f(t)\;.\label{v2mc2}
    \eea
\subsection{Class 3}
In this class we take $\bL=\bL_3 \equiv x\ro_t+t \ro_x$. With this
choice for $\bL$, \eqs{e1} and \e{e0} take the following form
    \bea
    &&Px=-2\ro_t L_0\;, \label{em1}\\
    &&Pt=2\ro_x L_0\label{em2}\\
    &&P L_0=-\ro_t^2 L_0 +\ro_x^2 L_0+x\ro_t V_1 +t\ro_x V_1\;,\label{em3}
    \eea
These equations are easily solved with the following change of
variables
    \be
    u=\frac{t}{x}\;,v=t^2-x^2\;. \label{changem3}
    \ee
The  general solutions are
    \bea
    &&L_0=L_0(u)\;, \label{l0mc3}\\
    &&P=-\frac{2}{x^2} L'_0\label{pmc3}\\
    &&V_1(u,v)=-\frac{1}{v}\left(L_0^2(u)+(1-u^2)L'_0(u)+f(v)\right)\;,\label{v1mc3}\\
    &&V_2(u,v)=-\frac{1}{v}\left(L_0^2(u)-(1-u^2)L'_0(u)+f(v)\right)\;.\label{v2mc3}
    \eea
\section{Poincar{\' e} half plane $(AdS_2)$}
Poincar{\' e} half plane is a 2-dimensional Riemannian manifold
with constant Gaussian curvature $\kappa=-1$. In fact it is  a
space with coordinates $(x,y)$ with $y>0$ and its metric is given
by $g_{ab}=diag(1/y^2,1/y^2)$.

In this space \eq{e2} takes the following form
   \bea
    &&L^y=y\ro_x L^x\;, \label{p1}\\
    &&L^y=y\ro_y L^y\;, \label{p2}\\
    &&\ro_x L^y +\ro_y L^x=0\;, \label{p3}
    %&&2y^2\ro_x L_0=P L^x\;, \label{p4}\\
    %&&2y^2\ro_y L_0=P L^y\;, \label{p5}\\
    %&&P L_0=y^2(\ro_x^2 L_0 +\ro_y^2 L_0)+L^x\ro_x V_1 +L^y\ro_y V_1\;, \label{p6}
    \eea
It can be easily shown that the most general solution of \eqs{p1} -- \e{p3}  for
$\bL=L^x\ro_x+L^y\ro_y$ is of the form of \eq{l} with
    \bea
    &&\bL_1=\ro_x\;,\label{lp1}\\
    &&\bL_2=x\ro_x+y\ro_y\;, \label{lp2}\\
    &&\bL_3=(x^2-y^2)\ro_x+2xy\ro_y\;. \label{lp3}
    \eea
These are in fact Killing vector fields of Poincare half plane.
They satisfy the algebra of  isometry group of this space namely
$SL(2, \R)$.
    \be
    [\bL_1,\bL_2]=\bL_1\;,[\bL_2,\bL_3]=\bL_3\;,[\bL_1,\bL_3]=2\bL_2\;.
    \label{sl2r}
    \ee
Again we distinguish three classes and  investigate each class
separately
\subsection{Class 1}
In this case we take $\bL=\bL_1\equiv \ro_x$. Then \eqs{e1} and \e{e0} simplify as
follows
    \bea
    &&P=2y^2\ro_x L_0\;, \label{pc1e1}\\
    &&\ro_y L_0 =0 \;, \label{pc1e2}\\
    && PL_0=y^2(\ro_x^2 L_0 +\ro_y^2 L_0)+\ro_x V_1\;. \label{pc1e3}
    \eea
From \eq{pc1e2} one can easily see that $L_0$ is a function of $x$ only and the
solution for $P$, $V_1$ and $V_2$ are as the following
    \bea
    &&L_0=L_0(x)\;, \label{pc1l0}\\
    &&P=2y^2 L'_0\label{pc1p}\\
    &&V_1(x,y)=y^2\left(L_0^2(x)-L'_0(x)+f(y)\right)\;,\label{pc1v1}\\
    &&V_2(x,y)=y^2\left(L_0^2(x)+L'_0(x)+f(y)\right)\;,\label{pc1v2}
    \eea
\subsection{Class 2}
This class is given by $\bL=\bL_2\equiv x\ro_x+y\ro_y$. Then \eqs{e1} and \e{e0} take
the following form
    \bea
    &&Px=2y^2\ro_x L_0 \;, \label{pc2e1}\\
    &&Py=2y^2\ro_y L_0 \;, \label{pc2e2}\\
    && PL_0=y^2(\ro_x^2 L_0 +\ro_y^2 L_0)+x \ro_x V_1+ y\ro_y V_1\;, \label{pc2e3}
    \eea
One can easily solve these equations with the following change of variables
    \be
    u=x^2+y^2\;\;, v=\frac{x}{y}\;. \label{changep2}
    \ee
The solutions are
    \bea
    &&L_0=L_0(u)\;, \label{pc2l0}\\
    &&P=4y^2 L'_0\label{pc2p}\\
    &&V_1(u,v)=\frac{1}{1+v^2}(L_0^2(u)-2uL'_0(u)+f(v))\;,\label{pc2v1}\\
    &&V_2(u,v)=\frac{1}{1+v^2}(L_0^2(u)+2uL'_0(u)+f(v))\;,\label{pc2v2}
    \eea
Here again one can notice to the relationship between the new
variables and the integral curves of the Killing vector field. In
fact the family of curves given by $v={\rm const}$ is the family
of integral curves of the Killing vector field $\bL_2$. These are
normal to the family of curves given by $u={\rm const.}$
\subsection{Class 3}
In this case we take $\bL=\bL_3\equiv (x^2-y^2)\ro_x + 2xy\ro_y$.
Then \eqs{e1} and \e{e0} read
    \bea
    &&P(x^2-y^2)=2y^2\ro_x L_0 \;, \label{pc3e1}\\
    &&P(2xy)=2y^2\ro_y L_0 \;, \label{pc3e2}\\
    && PL_0=y^2(\ro_x^2 L_0 +\ro_y^2 L_0)+(x^2-y^2) \ro_x V_1+ 2xy\ro_y V_1\;, \label{pc3e3}
    \eea
One can easily solve these equations with the following change of variables
    \be
    u=\frac{x^2+y^2}{x}\;\;, v=\frac{x^2+y^2}{y}\;. \label{changep3}
    \ee
The solutions are
    \bea
    &&L_0=L_0(u)\;, \label{pc3l0}\\
    &&P=\frac{2y^2}{x^2} L'_0\label{pc3p}\\
    &&V_1(u,v)=\frac{1}{v^2}(L_0^2(u)-u^2 L'_0(u)+f(v))\;,\label{pc3v1}\\
    &&V_2(u,v)=\frac{1}{v^2}(L_0^2(u)+u^2 L'_0(u)+f(v))\;,\label{pc3v2}
    \eea
\section{de Sitter Plane $(dS_2)$}
For de Sitter plane we choose the notation $x^1=t$ and $x^2=x$
for coordinates. The  metric is given by
$g_{ab}=diag(-1/t^2,1/t^2)$.

In this space \eq{e2} takes the following form
   \bea
    &&L^t=t\ro_x L^x\;, \label{d1}\\
    &&L^t=t\ro_t L^t\;, \label{d2}\\
    &&\ro_x L^t -\ro_t L^x=0\;, \label{d3}
    %&&2y^2\ro_x L_0=P L^x\;, \label{p4}\\
    %&&2y^2\ro_y L_0=P L^y\;, \label{p5}\\
    %&&P L_0=y^2(\ro_x^2 L_0 +\ro_y^2 L_0)+L^x\ro_x V_1 +L^y\ro_y V_1\;, \label{p6}
    \eea
It can be easily shown that the most general solution of \eqs{d1}
-- \e{d3}  for $\bL=L^t\ro_t+L^x\ro_x$ is of the form of \eq{l}
with
    \bea
    &&\bL_1=\ro_x\;,\label{ld1}\\
    &&\bL_2=x\ro_x+t\ro_t\;, \label{ld2}\\
    &&\bL_3=(x^2+t^2)\ro_x+2xt\ro_t\;. \label{ld3}
    \eea
These are in fact Killing vector fields of de Sitter  plane. They
satisfy the algebra of  isometry group of this space namely
$SL(2, \R)$.
    \be
    [\bL_1,\bL_2]=\bL_1\;,[\bL_2,\bL_3]=\bL_3\;,[\bL_1,\bL_3]=2\bL_2\;.
    \label{sl2rd}
    \ee
Again we distinguish three classes and  investigate each class
separately
\subsection{Class 1}
In this case we take $\bL=\bL_1\equiv \ro_x$. Then \eqs{e1} and
\e{e0} simplify as follows
    \bea
    &&P=2t^2\ro_x L_0\;, \label{d1e1}\\
    &&\ro_t L_0 =0 \;, \label{d1e2}\\
    && PL_0=t^2(\ro_x^2 L_0 -\ro_t^2 L_0)+\ro_x V_1\;. \label{d1e3}
    \eea
From \eq{d1e2} one can easily see that $L_0$ is a function of $x$
only and the solution for $P$, $V_1$ and $V_2$ are as the
following
    \bea
    &&L_0=L_0(x)\;, \label{d1l0}\\
    &&P=2t^2 L'_0\label{d1p}\\
    &&V_1(t,x)=t^2\left(L_0^2(x)-L'_0(x)+f(t)\right)\;,\label{d1v1}\\
    &&V_2(t,x)=t^2\left(L_0^2(x)+L'_0(x)+f(t)\right)\;,\label{d1v2}
    \eea
\subsection{Class 2}
This class is given by $\bL=\bL_2\equiv x\ro_x+t\ro_t$. Then
\eqs{e1} and \e{e0} take the following form
    \bea
    &&Px=2t^2\ro_x L_0 \;, \label{d2e1}\\
    &&Pt=-2t^2\ro_t L_0 \;, \label{d2e2}\\
    && PL_0=t^2(\ro_x^2 L_0 -\ro_t^2 L_0)+x \ro_x V_1+ t\ro_t V_1\;, \label{d2e3}
    \eea
One can easily solve these equations with the following change of
variables
    \be
    u=x^2-t^2\;\;, v=\frac{x}{t}\;. \label{changed2}
    \ee
The solutions are
    \bea
    &&L_0=L_0(u)\;, \label{d2l0}\\
    &&P=4t^2 L'_0\label{d2p}\\
    &&V_1(u,v)=\frac{1}{v^2-1}(L_0^2(u)-2uL'_0(u)+f(v))\;,\label{d2v1}\\
    &&V_2(u,v)=\frac{1}{v^2-1}(L_0^2(u)+2uL'_0(u)+f(v))\;,\label{d2v2}
    \eea
Here again one can notice to the relationship between the new
variables and the integral curves of the Killing vector field. In
fact the family of curves given by $v={\rm const}$ is the family
of integral curves of the Killing vector field $\bL_2$. These are
normal to the family of curves given by $u={\rm const.}$
\subsection{Class 3}
In this case we take $\bL=\bL_3\equiv (x^2+t^2)\ro_x + 2xt\ro_t$.
Then \eqs{e1} and \e{e0} read
    \bea
    &&P(x^2+t^2)=2t^2\ro_x L_0 \;, \label{d3e1}\\
    &&P(2xt)=-2t^2\ro_t L_0 \;, \label{d3e2}\\
    && PL_0=t^2(\ro_x^2 L_0 -\ro_t^2 L_0)+(x^2+t^2) \ro_x V_1+ 2xt\ro_t V_1\;, \label{d3e3}
    \eea
One can easily solve these equations with the following change of
variables
    \be
    u=\frac{x^2-t^2}{x}\;\;, v=\frac{x^2-t^2}{t}\;. \label{changed3}
    \ee
The solutions are
    \bea
    &&L_0=L_0(u)\;, \label{d3l0}\\
    &&P=\frac{2t^2}{x^2} L'_0\label{d3p}\\
    &&V_1(u,v)=\frac{1}{v^2}(L_0^2(u)-u^2 L'_0(u)+f(v))\;,\label{d3v1}\\
    &&V_2(u,v)=\frac{1}{v^2}(L_0^2(u)+u^2 L'_0(u)+f(v))\;,\label{d3v2}
    \eea
\section{Sphere}
For a sphere with unit radius we take the coordinates as $x^1=\theta$ and $x^2=\phi$.
The metric is $g_{ab}=diag(1,\sin^2 \theta)$. Then \eq{e2} implies
    \bea
    &&\ro_\theta L^\theta=0\;, \label{se1}\\
    &&\tan \theta \ro_\phi L^\phi+L^\theta=0 \;, \label{se2}\\
    &&\ro_\phi L^\theta +\sin^2\theta \ro_\theta L^\phi=0\;.\label{se3}
    \eea
The general solution of the above equations, $\bL=L^\theta
\ro_\theta +L^\phi \ro_\phi$, is given by \eq{l} with
    \bea
    &&\bL_1=\ro_\phi\;,\label{l1s}\\
    &&\bL_2=\cos \phi \ro_ \theta -\sin \phi \cot \theta \ro_\phi\;, \label{l2s}\\
    &&\bL_3=\sin \phi \ro_ \theta +\cos \phi \cot \theta \ro_\phi\;. \label{l3s}
    \eea
Here again we obtained the Killing vector fields of $S^2$ (two dimensional sphere),
which satisfy the algebra of isometry group of $S^2$, namely $SO(3)$
    \be
    [\bL_1,\bL_3]=\bL_2\;\;,[\bL_3,\bL_2]=\bL_1\;\;,[\bL_2,\bL_1]=\bL_3\;,
    \label{s2algebra}
    \ee
Again we investigate each of the above solutions in a separate
class. Here again one should note that there are only one class
from a physical point of view. In fact all three classes given
below are equivalent. But algebraically one can use these classes
to construct the most general intertwining operator.
\subsection{Class 1}
In this class $\bL=\bL_1\equiv \ro_\phi$. Then \eqs{e1} , \e{e0} read
    \bea
    &&\ro_\theta L_0 =0\;, \label{sc1e1}\\
    &&P \sin^2 \theta - 2\ro_\phi L_0 =0\;, \label{sc1e2}\\
    &&P L_0=\cot \theta \ro_\theta L_0 +\ro_\theta^2 L_0 +\frac{1}{\sin^2
    \theta}\ro^2_\phi L_0 +\ro_\phi V_1\;. \label{sc1e3}
    \eea
The above equations result in the following general solutions
    \bea
    &&L_0=L_0(\phi)\;, \label{sc1l0}\\
    &&P=\frac{2}{\sin^2 \theta} L'_0\label{sc1p}\\
    &&V_1(\theta,\phi)=\frac{1}{\sin^2 \theta}(L_0^2- L'_0+f(\theta))\;,\label{sc1v1}\\
    &&V_2(\theta,\phi)=\frac{1}{\sin^2 \theta}(L_0^2+ L'_0+f(\theta))\;.\label{sc1v2}
    \eea
\subsection{Class 2}
In this class $\bL=\bL_2\equiv \cos \phi \ro_ \theta -\sin \phi \cot \theta \ro_\phi$.
Then \eqs{e1} , \e{e0} read
    \bea
    &&P \cos \phi=2\ro_\theta L_0 \;, \label{sc2e1}\\
    &&P \sin \theta \cos \theta \sin \phi + 2\ro_\phi L_0 =0\;, \label{sc2e2}\\
    &&P L_0=\cot \theta \ro_\theta L_0 +\ro_\theta^2 L_0 +\frac{1}{\sin^2
    \theta}\ro^2_\phi L_0 +\cos \phi \ro_\theta V_1-\cot \theta \sin \phi
    \ro_\phi V_1\;. \label{sc2e3}
    \eea
One can easily solve these equations with the help of the
following change of variables
    \be
    u=\cos \phi \tan \theta\;\;, v=\sin \phi \sin \theta\;. \label{changes2}
    \ee
The  general solutions are given below
    \bea
    &&L_0=L_0(u)\;, \label{sc2l0}\\
    &&P=\frac{2}{\cos^2 \theta} L'_0\label{sc2p}\\
    &&V_1(u,v)=\frac{1}{1-v^2}(L_0^2(u)-(1+u^2) L'_0(u)+f(v))\;,\label{sc2v1}\\
    &&V_2(u,v)=\frac{1}{1-v^2}(L_0^2(u)+(1+u^2) L'_0(u)+f(v))\;,\label{sc2v2}
    \eea
where $f(v)$ is an arbitrary function of $v$.
\subsection{Class 3} In this class $\bL=\bL_3\equiv \sin \phi \ro_ \theta +\cos \phi
\cot \theta \ro_\phi$. Then \eqs{e1} , \e{e0} read
    \bea
    &&P \sin \phi=2\ro_\theta L_0 \;, \label{sc3e1}\\
    &&P \sin \theta \cos \theta \cos \phi - 2\ro_\phi L_0 =0\;, \label{sc3e2}\\
    &&P L_0=\cot \theta \ro_\theta L_0 +\ro_\theta^2 L_0 +\frac{1}{\sin^2
    \theta}\ro^2_\phi L_0 +\sin \phi \ro_\theta V_1+\cot \theta \cos \phi
    \ro_\phi V_1\;. \label{sc3e3}
    \eea
One can easily solve these equations with the help of following change of variables
    \be
    u=\sin \phi \tan \theta\;\;, v=\cos \phi \sin \theta\;. \label{changes3}
    \ee
The  general solution are given below
    \bea
    &&L_0=L_0(u)\;, \label{sc3l0}\\
    &&P=\frac{2}{\cos^2 \theta} L'_0\label{sc3p}\\
    &&V_1(u,v)=\frac{1}{1-v^2}(L_0^2(u)-(1+u^2) L'_0(u)+f(v))\;,\label{sc3v1}\\
    &&V_2(u,v)=\frac{1}{1-v^2}(L_0^2(u)+(1+u^2) L'_0(u)+f(v))\;,\label{sc3v2}
    \eea
where $f(v)$ is an arbitrary function of $v$.
\section{Torus}
Consider a torus which it's minor and major radii are $1$ and $R$
respectively. The coordinates are $x^1=\theta$ and $x^2=\phi$
which show the position of minor and major radii respectively. The
metric is $g_{ab}=diag\left(1,(R+\cos \theta)^2\right)$. \eq{e2}
reads
    \bea
    &&\ro_\theta L^\theta=0\;, \label{lt1}\\
    &&(R+\cos \theta)^2\ro_\theta L^\phi-\ro_\phi L^\theta=0\;, \label{lt2}\\
    &&(R+\cos \theta)\ro_\phi L^\phi -\sin \theta L^\theta=0\;, \label{lt3}
    \eea
The only solution of the above equations is $\bL=\ro_\phi$. In
fact this is the only vector field on the torus. The isometry
group corresponding to this vector field is $U(1)$. Then one can
easily find the solutions for $L_0$, $P$, $V_1$ and $V_2$,
    \bea
    &&L_0=L_0(\phi)\;, \label{tl0}\\
    &&P=\frac{2}{(R+\cos \theta)^2} L'_0\label{tp}\\
    &&V_1(\theta,\phi)=\frac{1}{(R+\cos \theta)^2}(L_0^2- L'_0+f(\theta))\;,\label{tv1}\\
    &&V_2(\theta,\phi)=\frac{1}{(R+\cos \theta)^2}(L_0^2+ L'_0+f(\theta))\;,\label{tv2}
    \eea
where $f(\theta)$ is an arbitrary function of $\theta$.
\section{Applications}
In this section we give two examples for the formalism given in
the above sections. The first example is about quantum systems
with equispaced energy levels and the second one is quantum
systems which their spectrum is like the spectrum of a free
particle.
\subsection{Equispaced Energy Levels}
Quantum systems with equispaced energy levels are of some
importance in condensed matter Physics and optics
{\cite{equispaced}}. In the formalism given in this paper it it
is easy to investigate such systems. In fact if $P=V_2-V_1$ be a
nonzero constant, then both the Hamiltonians $H_1$ and $H_2$ have
equispaced energy levels and the operator $\LL$ will be the
lowering operator for the Hamiltonian $H_1$, i.e.
$[\LL,H_1]=P\LL$. This means that $P$ is the space between energy
levels. If we have a look at the solutions on surfaces
investigated in this paper we will see that among these surfaces
only the Euclidean plane allows for such solutions i.e. for
solutions in which $P$ is constant and consequently the energy
levels are equispaced. A question arises naturally: Is the
Euclidean plane the only surface that allows for the systems with
equispaced energy levels? If not, then  what is the general
solution? To answer this question we will put $P=const.$ in
\eqs{e2} -- \e{e0} and see what conditioins this choice will put
on the surface. One can put $P=1$ without loss of generality.
This means that we measure energy in $P$ units. With this choice
and using \eqs{e2} and \e{e1} one can easily verify that the
Killing vector field must satisfy the following condition
    \be
    \nabla_a L^b\equiv \ro_a L_b +\Gamma^b_{ac} L^c=0\;, a,b=1,2
    \label{condition}
    \ee
where $\nabla_a L^b$ is the covariant derivative of $L^b$ and
$\Gamma^b_{ac}:=\frac{1}{2} g^{bd}\left[\ro_a g_{dc} +\ro_c
g_{ad} -\ro_d g_{ac}\right]$ is the connection. Now using the
definition of curvature tensor
$\left[\nabla_a,\nabla_b\right]L^c={R_{bad}}^cL^d$ and
\eq{condition} one can easily verify that \eq{condition} is
satisfied only if the curvature tensor equals to zero
identically. The curvature tensor in two dimensions has just one
independent component and it is $R_{1212}=\frac{1}{2}Rg$ where $R$
is the curvature of the surface and $g$ is the determinant of the
metric tensor. This fact together with the definition of
curvature tensor and \eq{condition} results in
    \bea
    &&\frac{1}{2}R g(g^{12}L^1-g^{11}L^2)=0\label{flat1}\;,\\
    &&\frac{1}{2}R g(g^{22}L^1-g^{21}L^2)=0\label{flat2}\;.
    \eea
These equations are consistent only if $R=0$. Therefore in our
formalism systems with equispaced energy levels can only exist on
flat surfaces which are surfaces with zero curvature tensor. This
means that for example on a compact surface like torus or sphere
one can not construct a quantum system with equispaced energy
levels. This is also verified from the fact that the condition
$P=const.$ leads to $\nabla^2 L_0=0$. Because this is the Laplace
equation which has a trivial solution on a compact surface.
\subsection{Free Like Particles}
By {\it free like} we mean a particle which moves in a nontrivial
potential, but its spectrum is essentially the free particle
spectrum. This makes sense only on a compact surface because on a
compact surface the free particle spectrum is discrete.  In our
formalism this potential is the partner of constant potential.
For a constant potential $V_1$ \eq{e0} leads to
    \be
    PL_0=\nabla^2L_0\;.
    \label{constant}
    \ee
Using the above equation and \eq{e1} one can find $L_0$ and $P$.
We construct the potentials of free like particles on sphere and
torus.

\noindent {\bf \large Sphere:} To find free like potential on a
sphere one can easily put $f(\theta)=1$  and $L_0(\phi)=\tan
\phi$ in \eqs{sc1v1} and \e{sc1v2}. Then one gets
$V_1(\theta,\phi)=0$ and
    \be
    V_2(\theta,\phi)=\frac{2}{\sin^2\theta \cos^2 \phi}\;.
    \label{sfree}
    \ee
\noindent {\bf \large Torus:} On a torus using \eqs{tv1} and
\e{tv2} and putting $f(\theta)=1$  and $L_0(\phi)=\tan \phi$ one
arrives at $V_1(\theta, \phi)=0$ and
    \be
    V_2(\theta,\phi)=\frac{2}{(\cos \theta+R)^2 \cos^2 \phi}\;.
    \label{tfree}
    \ee

\section{Summary and Concluding Remarks}
In this paper we investigated the intertwined Hamiltonians in
some two dimensional curved spaces. We found the general forms of
intertwined potentials and the intertwining operators. It was
shown that the intertwining operators are closely related to the
Killing vector fields of the corresponding space. In fact the
intertwining operator is the Killing vector field plus a real
valued function in the corresponding space. This real valued
function is closely connected to the integral curves of
corresponding Killing vector filed.

A comment on the case of $P\equiv V_2-V_1=0$ is in order. In this
case the intertwining operator $\LL$ commutes with $H_1=H_2$. In
this case, $L_0$ is a constant and the potential $V_1=V_2$ is a
function of the variable `$v$' only. As $v=const.$ is an integral
curve of the Killing vector field, this means that in quantum
systems in which the potential is constant along the integral
curves of the Killing vector field, the corresponding Killing
vector field is a constant of motion.

Another interesting case is the the case in which $P\equiv
V_2-V_1=const.$ In this case the energy levels of the Hamiltonian
$H_1$ (and $H_2$) are equispaced. It can be easily verified that
among the spaces studied in this paper, this happens only in
Euclidean plane (cases 1 and 2). We have shown that only surfaces
with zero curvature allow for such solutions.

\section*{Acknowledgment}
Financial supports of Isfahan University of Technology is acknowledged.

\end{document}